\documentstyle[twoside,fleqn,espcrc2,epsf]{article}
\newcommand{\bc}{\begin{center}}
\newcommand{\ec}{\end{center}}
\newcommand{\beqn}{\begin{equation}}
\newcommand{\eeqn}{\end{equation}}
\newcommand{\barr}{\begin{eqnarray}}
\newcommand{\earr}{\end{eqnarray}}

\def\del{\partial}

\def\eg {{\it e.g}. }

\def\tr {\mbox{tr}}

\def\PL #1 #2 #3 {Phys. Lett. {\bf#1} (#2) #3}
\def\NP #1 #2 #3 {Nucl. Phys. {\bf#1} (#2) #3}
\def\NPP #1 #2 #3 {Nucl. Phys. {\bf B} (Proc. Suppl.) {\bf#1} (#2) #3}
\def\ZP #1 #2 #3 {Z.~Phys. {\bf#1} (#2) #3}
\def\PR #1 #2 #3 {Phys. Rev. {\bf#1} (#2) #3}
\def\PP #1 #2 #3 {Phys. Rep.{\bf#1} (#2) #3}
\def\PRL #1 #2 #3 {Phys. Rev. Lett. {\bf#1} (#2) #3}
\def\PTP #1 #2 #3 {Prog. Theor. Phys. {\bf#1} (#2) #3}
\def\MPL #1 #2 #3 {Mod. Phys. Lett. {\bf#1} (#2) #3}
\def\IJM #1 #2 #3 {Int. J.~Mod. Phys. {\bf#1} (#2) #3}
\title{Existence of Chiral-Asymmetric Zero Modes in the Background
of QCD-Monopoles\thanks{Talk presented by S.~Sasaki at LATTICE 97, 
Edinburgh.}
}

\author{S. Sasaki\address{Yukawa Institute for Theoretical Physics
(YITP), Kyoto University, %\\
Kyoto 606-01, Japan}%
\thanks{Research Fellow of the Japan Society for the Promotion of Science.}
and 
O. Miyamura\address{Department of Physics, Hiroshima University, %\\
Higashi-Hiroshima 739, Japan}}
       
\begin{document}

\begin{abstract}
We study topological aspects of the QCD vacuum structure in
${\rm SU}(2)$ lattice gauge theory with the abelian gauge fixing.
The index of the Dirac operator is measured by using the Wilson 
fermion in the quenched approximation.
We find chiral-asymmetric zero modes in background fields dominated 
by QCD-monopoles without any cooling.
\end{abstract}

\maketitle

\section{Topological objects in the QCD vacuum}
There are two distinct pictures of the vacuum structure in QCD.
One is based on the appearance of color-magnetic monopoles (QCD-monopoles) 
after performing the abelian gauge fixing \cite{tHooft}.
The recent lattice QCD simulations show that the dual Meissner effect brought 
to the QCD vacuum by QCD-monopole condensation \cite{Rev1}. 
Hence, color confinement could be regarded 
as the dual version of the superconductivity. 
On the other hand, QCD has also classical and non-trivial gauge 
configurations (instantons) as topological defects in the Euclidean 4-space.
As well known, the instanton liquid characterized by a random ensemble of 
instantons and anti-instanton succeeds in 
explaining several properties of light hadrons, \eg 
spontaneous chiral-symmetry breaking (S$\chi$SB) \cite{Rev2}.

It seems that QCD-monopoles and instantons are relevant topological objects 
for the description of each phenomenon.
Here, we should mention that QCD-monopoles would play an essential role on
non-perturbative features of QCD, which includes S$\chi$SB.
This possibility was studied by using the Schwinger-Dyson equation
with the gluon propagator in the background of condensed monopoles 
\cite{Suganuma1}.
The idea of providing S$\chi$SB due to 
QCD-monopole condensation was supported by
the lattice simulations \cite{Miyamura1,Woloshyn}.
Thus, these results shed new light on the non-trivial relation  
between QCD-monopoles and instantons. Recently, both analytic and
numerical works have shown the existence of the strong correlation 
between these topological objects  \cite{Rev1}. 

To appreciate further justification for this relation,
one should be reminded of the Atiyah-Singer index theorem.
The index of the Dirac operator, which corresponds to the
number of chiral-asymmetric zero modes, is equal to the 
topological charge \cite{Rev2}. Thus, instantons can be regarded as 
important topological  
objects related to the ${\rm U_{A}}(1)$ anomaly \cite{Rev2}.
If the Atiyah-Singer index theorem holds in the background 
of QCD-monopoles \cite{Sasaki}, we would find the monopole
dominance for the ${\rm U_{A}}(1)$ anomaly. We then study 
the eigenvalue spectrum of the Dirac operator in background fields
dominated by QCD-monopoles in order to examine 
the existence of chiral-asymmetric zero modes \cite{Sasaki}.

\section{Monopole-dominating background field}
We generate gauge configurations by using the Monte Carlo
simulation on ${\rm SU}(2)$ lattice gauge theory 
with the standard Wilson action. 
The gauge transformation 
is actually carried out by 
maximizing the gauge dependent variable $R$;
%
% eq. 
%
\beqn
R=\sum_{n,\;\mu}\tr\left\{
\sigma_{3} U_{\mu}(n) \sigma_{3} U^{\dag}_{\mu}(n) \right\}\;\;.
\eeqn
This partial gauge fixing is called as 
the maximally abelian (MA) gauge \cite{Kronfeld}.
Once the abelian gauge fixing is done by this procedure, we  
factorize the SU(2) link variable $U_{\mu}$ into the 
${\rm U}(1)$ link variable $u_{\mu}$ and an off-diagonal part $M_{\mu}$ 
as $U_{\mu}(n)=M_{\mu}(n)\cdot u_{\mu}(n)$ where
$u_{\mu}(n)\equiv\exp \{ i\sigma_{3}\theta_{\mu}(n) \}$ and
$M_{\mu}(n)\equiv\exp \{ 
i\sigma_{1}C^{1}_{\mu}(n)+i\sigma_{2}C^{2}_{\mu}(n) \}$.
$\theta_{\mu}$ is the ${\rm U}(1)$ gauge field and
$C^{1}_{\mu}$ and $C^{2}_{\mu}$ correspond to 
charged matter fields under a residual ${\rm U}(1)$ 
gauge transformation \cite{Kronfeld}.

In order to look for magnetic monopoles in terms of
${\rm U}(1)$ variables, we consider the product of 
${\rm U}(1)$ link variables around an elementary plaquette,
$u_{\mu \nu}(n)=u_{\mu}(n) u_{\nu}(n+{\hat \mu})
u^{\dag}_{\mu}(n+{\hat \nu}) u^{\dag}_{\nu}(n)=e^{i\sigma_{3}
\theta_{\mu \nu}(n)}$ with the ${\rm U}(1)$ field strength 
$\theta_{\mu \nu}(n)\equiv
\theta_{\nu}(n+{\hat \mu})-\theta_{\nu}(n)-\theta_{\mu}(n+{\hat \nu})
+\theta_{\mu}(n)$ \cite{Kronfeld}.
It should be noted that the ${\rm U}(1)$ plaquette variable 
is a multiple valued 
function as the ${\rm U}(1)$ field strength due to the compactness of
the residual ${\rm U}(1)$ gauge group. 
Then we can divide the ${\rm U}(1)$ field strength into two parts as
$\theta_{\mu \nu}={\bar \theta}_{\mu \nu}+2\pi
N_{\mu \nu}$
where ${\bar \theta}_{\mu \nu}$ is the regular part defined in 
$-\pi <{\bar \theta}_{\mu \nu}\leq \pi$ and 
$N_{\mu \nu}\in {\bf Z}$ is the modulo 
$2\pi$ of $\theta_{\mu \nu}$ \cite{DeGrand}.
Here, it is known that the abelian dominance for 
the SU(2) link variable as $U_{\mu}\simeq 
u_{\mu}$ in the MA gauge \cite{Rev1}. 
In this sense, the Dirac string can be identified by
the DeGrand-Toussaint's definition in the compact ${\rm U}(1)$ 
lattice gauge theory \cite{DeGrand}.

The ${\rm U}(1)$ gauge field can be decomposed 
into the regular part $\theta^{\rm Ph}_{\mu}$ and
the singular part $\theta^{\rm Ds}_{\mu}$
as $\theta^{L}_{\mu}=\theta^{\rm Ph}_{\mu}+\theta^{\rm Ds}_{\mu}$
{\it in the Landau gauge} ($\del_{\mu}\theta^{L}_{\mu}=0$) \cite{Miyamura1}.
Two parts are respectively defined by
%
% eq. 
%
\beqn
\theta^{\rm Ph}_{\mu}(n) \equiv \sum_{m,\;\lambda}G(n-m)\del_{\lambda}
{\bar \theta}_{\lambda \mu}(m) \;\;,
\eeqn
%
%
% eq. 
%
\beqn
\theta^{\rm Ds}_{\mu}(n) \equiv 2\pi\sum_{m,\;\lambda}G(n-m)\del_{\lambda}
N_{\lambda \mu}(m) 
\eeqn
where $G(n-m)$ is the lattice Coulomb propagator.
It is worth mentioning that the singular part $\theta^{\rm Ds}_{\mu}$ 
keeps essential contributions to
confining features of the Polyakov loop and finite 
quark condensate \cite{Miyamura1}.

%%%%%%%
% Fig %
%%%%%%%
\begin{figure}[t]
\epsfxsize=6.5cm
\centerline{\epsfbox{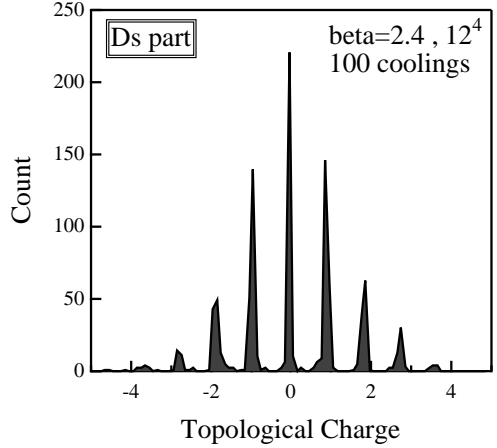}}
\caption{Histogram of topological charges 
in the background of $U^{\rm Ds}_{\mu}$
for 1000 configurations.}
\label{fig:topchr}
\end{figure}
%%%%%%%

Next, we define two types of background field, 
the photon-dominating (monopole-absent) part 
$U^{\rm Ph}_{\mu}$ and the monopole-dominating part $U^{\rm Ds}_{\mu}$
as 
%
% eq.
%
\beqn
U_{\mu}(n) = U^{\rm Ph}_{\mu}(n) \cdot u^{\rm Ds}_{\mu}(n) 
= U^{\rm Ds}_{\mu}(n) \cdot u^{\rm Ph}_{\mu}(n) 
\eeqn
where $u^{\;i}_{\mu}(n)\equiv
\exp\{i\sigma_{3}\theta^{\;i}_{\mu}(n)\}$ ($i=$ Ph or 
Ds) \cite{Miyamura2,Suganuma2}. 
It is noted that this definition corresponds to the reconstruction of 
the corresponding SU(2) variables 
from $u^{\;i}_{\mu}$ by multiplying the off-diagonal
factor in the Landau gauge ($\theta_{\mu} =
\theta_{\mu}^{L}+ \del_{\mu}\phi$) as
%
% eq. 
%
\beqn
{\tilde U}^{\;i}_{\mu}(n) \equiv {\tilde M}_{\mu}(n)\exp\{i\sigma_{3}
\theta^{\;i}_{\mu}(n)\} 
\eeqn
where ${\tilde U}^{\;i}_{\mu}(n)= d(n)U^{\;i}_{\mu}(n)d^{\dag}(n+\mu)$
and ${\tilde M}_{\mu}(n)=d(n)M_{\mu}(n)d^{\dag}(n)$  
with $d(n)=e^{i\phi(n)\sigma_{3}}$ \cite{Miyamura2,Suganuma2}. 

Here, it is worth measuring the topological charge in previously 
defined backgrounds.
As shown in Fig.\ref{fig:topchr}, we can observe the topological charge 
in the background of $U^{\rm Ds}_{\mu}$ \cite{Miyamura2,Suganuma2}.
However, non-zero value of the topological charge is not found 
in the background of $U^{\rm Ph}_{\mu}$, where instantons seem unable 
to live \cite{Miyamura2,Suganuma2}.

\section{Chiral-asymmetric zero modes}
In order to examine the eigenvalue of the Dirac operator 
$D \kern -2.2mm {/}$ on the lattice, we adopt the Wilson fermion 
operator \cite{Sasaki}. 
In the background of $U^{\;i}_{\mu}$, $D \kern -2.2mm {/}$
is expressed as
%
% eq. 
%
\barr
D \kern -2.2mm {/}({n,\;m}) \makebox[5.5cm]\nonumber \\
= \delta_{n,\;m}-\kappa \sum_{\mu} \left[
(1-\gamma_{\mu})U^{\;i}_{\mu}(n)\delta_{n+{\hat \mu},\;m} \right.
\makebox[0.5cm]\nonumber \\
\left.+(1+\gamma_{\mu})U^{\;i\;\dag}_{\mu}(n-{\hat \mu})
\delta_{n-{\hat \mu},\;m}\right]
\earr
where $\kappa$ is the hopping parameter.
The operator $D \kern -2.2mm {/}$ loses a feature as the hermitian operator 
owing to the discretization of the space-time. 
However, we can easily see that $\gamma_{5}{D \kern -2.2mm {/}}$
is a hermitian matrix. Then, the existence of chiral-asymmetric zero modes 
can be identified by the zero-line crossing in the eigenvalue spectrum 
of $\gamma_{5}{D \kern -2.2mm {/}}$ through 
the variation of $\kappa$ \cite{Itoh}.

\begin{figure}[hb]
\epsfxsize=6.4cm
\centerline{\epsfbox{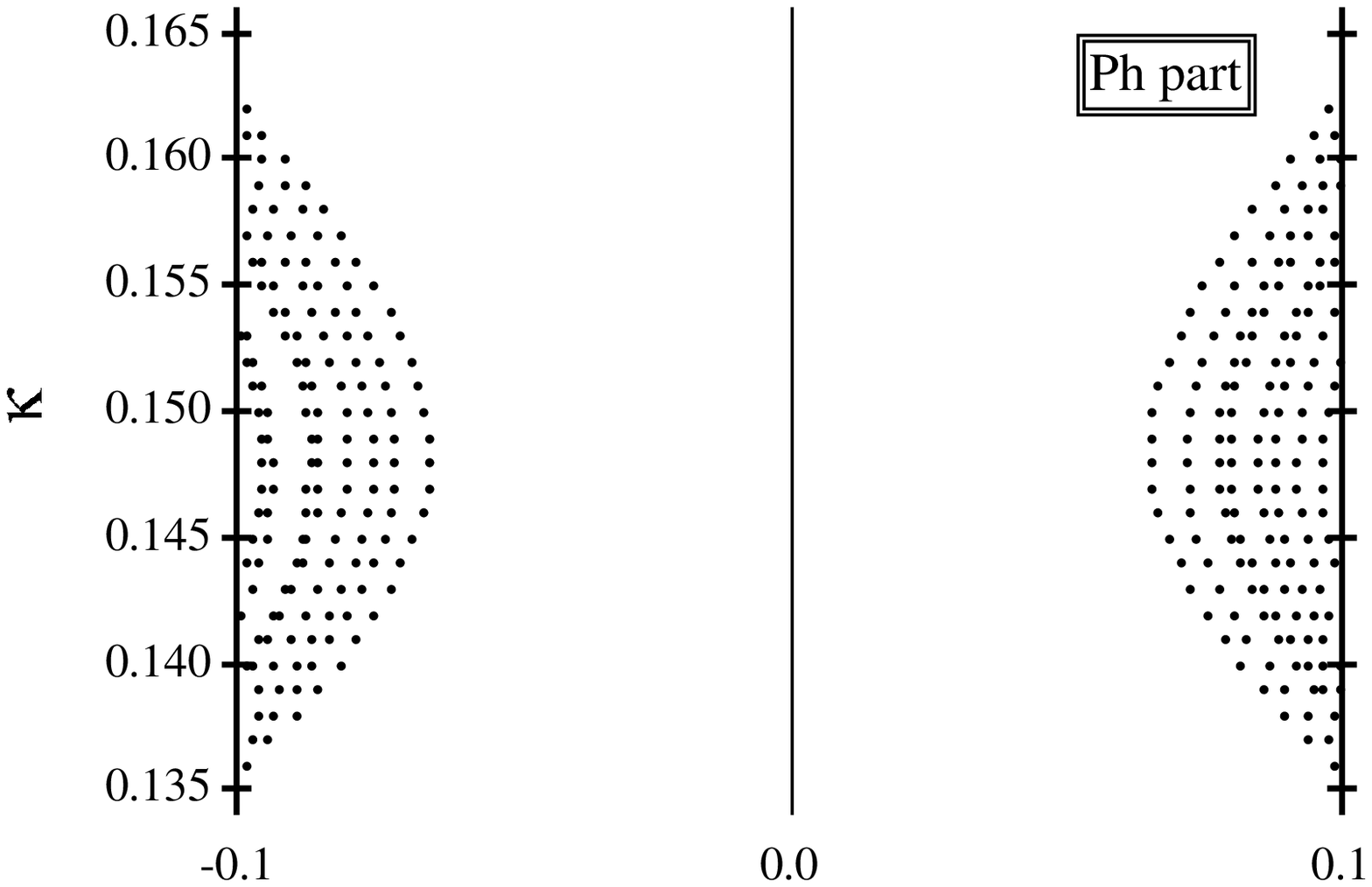}}

\vspace{0.3cm}
\epsfxsize=6.4cm
\centerline{\epsfbox{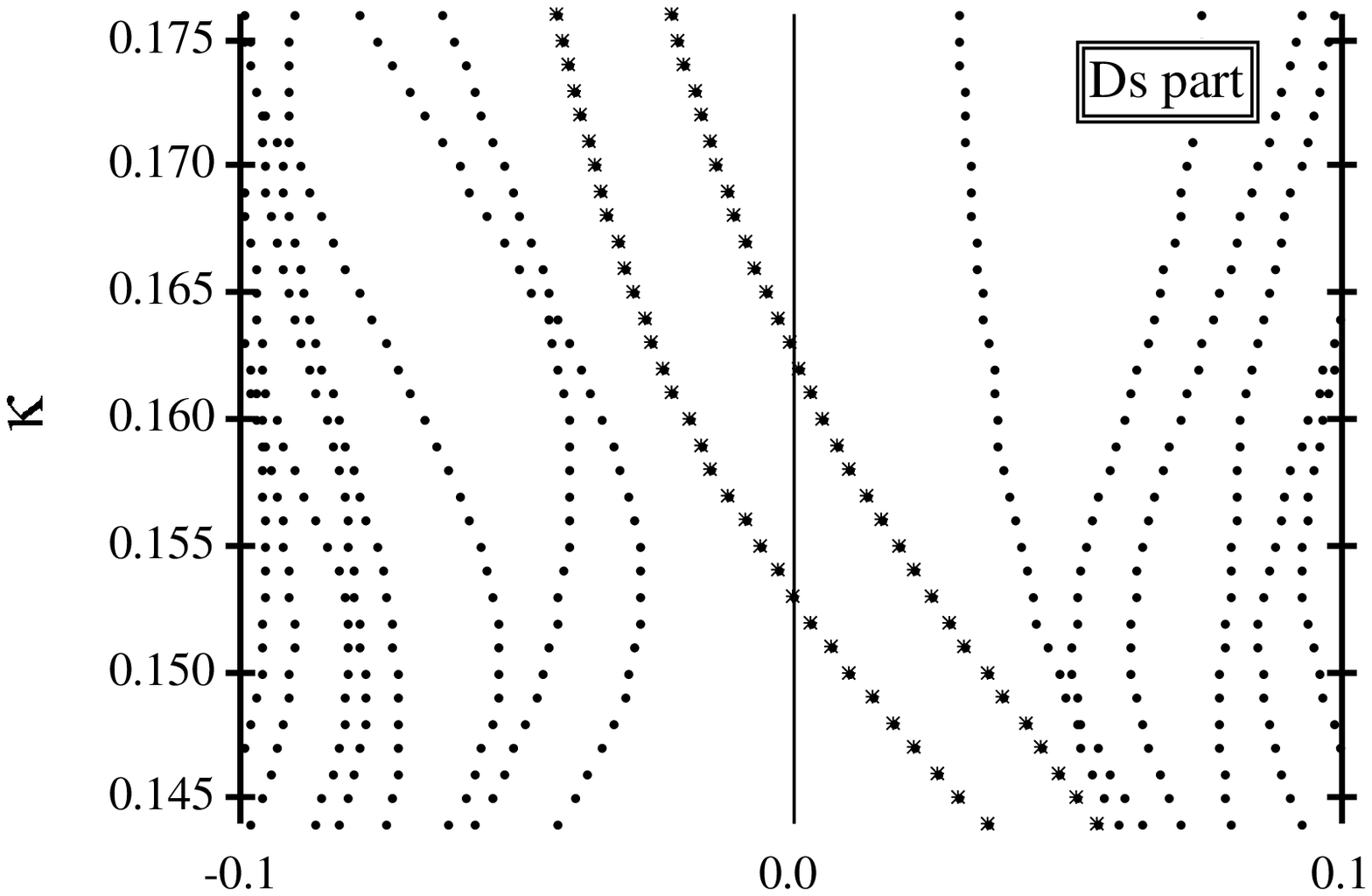}}

\vspace{0.3cm}
\epsfxsize=6.4cm
\centerline{\epsfbox{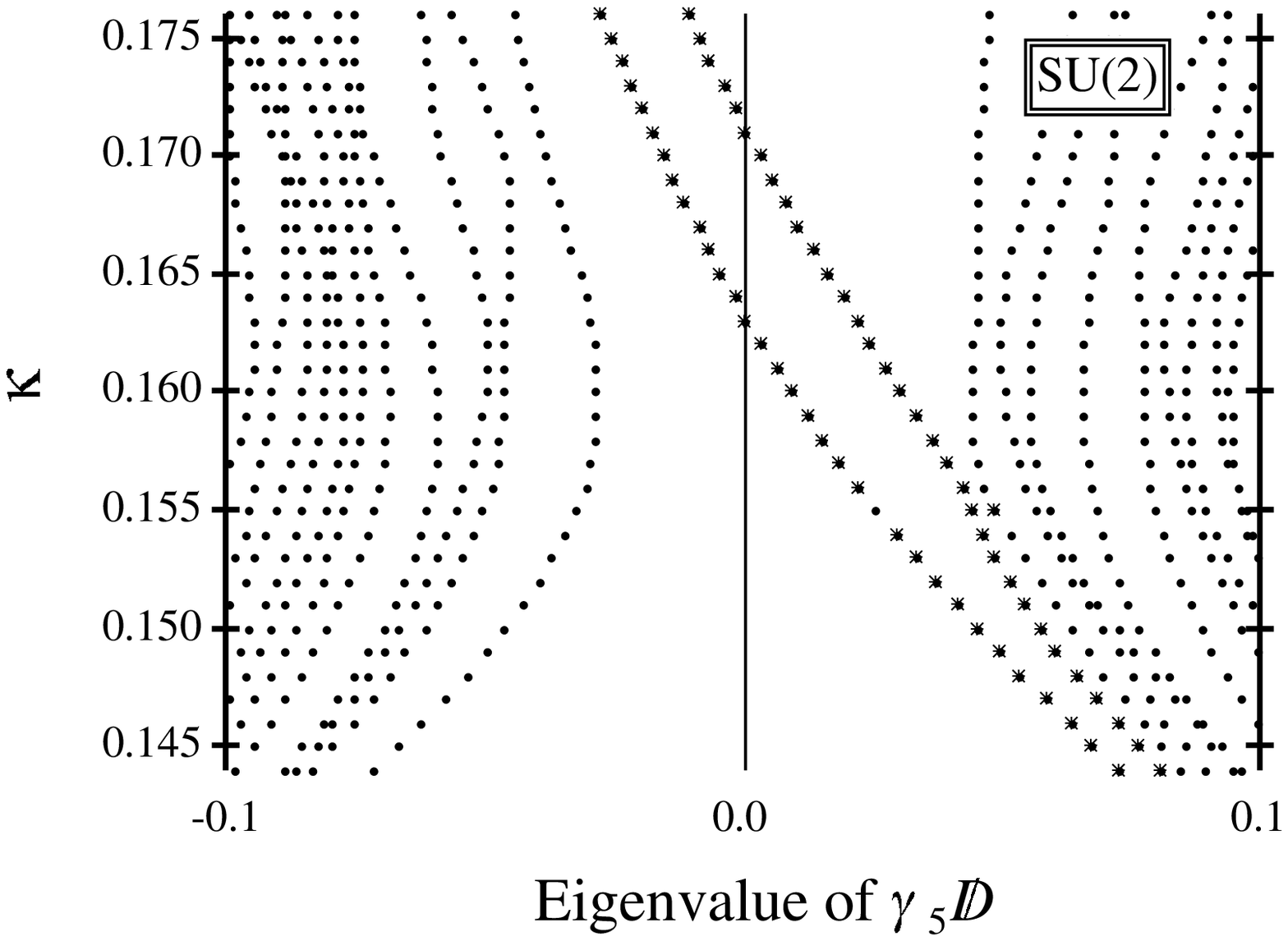}}
\caption{Typical examples of the low-lying spectra of $\gamma_{5}{D \kern -2.2mm {/}}$
in the background of (a) $U^{\rm Ph}_{\mu}$, (b) $U^{\rm Ds}_{\mu}$ and 
(c) $U_{\mu}$.}
\label{fig:Zero}
\end{figure}

We measure the eigenvalue of the 
operator $\gamma_{5}{D \kern -2.2mm {/}}$ 
in the background of $U^{\;i}_{\mu}$ ($i=$ Ph or Ds) and also 
in the original gauge field $U_{\mu}$ for 32 independent 
configurations {\it without any cooling}.
Fig.\ref{fig:Zero}(a)-(c) show the low-lying spectra in each background, 
which is based on the same gauge configuration, as typical examples.
We can see that 2 zero modes exist
in the monopole dominating background (Ds part) and 
the original ${\rm SU}(2)$ gauge field.
This remarkable coincidence for the number of the zero modes is not well 
identified in 6 configurations, but is confirmed in the rest 26 configurations. 
On the other hand, we can never find the corresponding zero modes in the 
monopole-absent background (Ph part) within 32 configurations.

In conclusion,
we have investigated the eigenvalue spectrum for the Dirac
operator in the background of the monopole-absent (Ph) part,
the monopole-dominating (Ds) part and the original gauge field
on an $8^{4}$ lattice at $\beta=2.4$ by using the Lanczos algorithm.
We have found the existence of chiral-asymmetric zero modes
in the background of QCD-monopoles, where instantons could survive.
%------------References--------------
%

%%%%%%%
% Fig %
%%%%%%%
%%%%%%%


\begin{thebibliography}{9}

\bibitem{tHooft} G.~'t~Hooft, \NP {B190} {1081} {455}.

\bibitem{Rev1} For a recent review article, see M.I.~Polikarpov,
Nucl. Phys. {\bf B} (Proc. Suppl.) {\bf 53} (1997)
134 and references therein.

\bibitem{Rev2} For a recent review article, see T.~Sch{\"a}fer and 
E.V.~Shuryak, hep-ph/9610451.

\bibitem{Suganuma1}
H.~Suganuma, S.~Sasaki and H.~Toki, \NP {B435} {1995} {207}.

\bibitem{Miyamura1}
O.~Miyamura, \PL {B353} {1995} {91}.

\bibitem{Woloshyn}
R.M.~Woloshyn, \PR {D51} {1995} {6411}.

\bibitem{Sasaki}
S.~Sasaki and O.~Miyamura, preprint YITP-97-35; hep-lat/9706001.

\bibitem{Kronfeld} 
A.G.~Kronfeld, M.L.~Laursen, G.~Schierholz and U.-J.~Wiese, 
\PL {B198} {1987} {516}.

\bibitem{Miyamura2}
O.~Miyamura and S.~Origuchi, in {\it Confinement 95} (World 
Scientific, 1995) 235.

\bibitem{Suganuma2}
H.~Suganuma, A.~Tanaka, S.~Sasaki and O.~Miyamura,
\NPP {47} {1996} {302}.

\bibitem{DeGrand} 
T.A.~DeGrand and D.~Toussaint, \PR {D22} {1980} {2478}.

\bibitem{Itoh} S.~Itoh, Y.~Iwasaki and T.~Yoshi{\'e}, \PR {D36} 
{1987} {527}.
%}
\end{thebibliography}
\end{document}